\documentclass[conference]{IEEEtran}
\IEEEoverridecommandlockouts

\usepackage[T1]{fontenc}
\usepackage[utf8]{inputenc}
\usepackage[english,turkish,main=turkish,shorthands=off]{babel}
\usepackage{amsmath}
\usepackage{amssymb}
\usepackage{graphicx}
\usepackage{booktabs}
\usepackage{cite}

\begin{document}

\makeatletter
\let\SAVEDmaketitle\maketitle
\let\SAVEDatmaketitle\@maketitle
\makeatother

\selectlanguage{english}
\renewcommand{\figurename}{Fig.}
\renewcommand{\tablename}{TABLE}

\title{Self-Attention Transformer-Based Detector\\
for Faster-than-Nyquist Signaling}

\author{\IEEEauthorblockN{Nurettin \c{S}afak, Osman Tokluoglu, Enver Cavus}
\IEEEauthorblockA{Department of Electrical and Electronics Engineering,
Ankara Y{\i}ld{\i}r{\i}m Beyaz{\i}t University, Ankara, T\"{u}rkiye\\
nsafaked@gmail.com, otokluoglu@aybu.edu.tr, ecavus@aybu.edu.tr}
}

\IEEEpubid{\makebox[\columnwidth]{\textbf{979-8-3195-1046-4/26/\$31.00 ©2026 IEEE}\hfill}
\hspace{\columnsep}\makebox[\columnwidth]{}}

\maketitle

\renewcommand{\abstractname}{Abstract}
\begin{abstract}
In this study, a novel encoder-only Transformer-based receiver architecture is
presented for BPSK signals transmitted over Faster-than-Nyquist (FTN) signaling
channels that introduce intentional inter-symbol interference (ISI) with a
compression factor of $\tau{=}0.8$. A complete end-to-end communication chain
encompassing BPSK modulation, RRC pulse shaping, and the ISI coefficients
arising from matched filtering was constructed and evaluated. The proposed
Transformer receiver was benchmarked against the optimal BCJR detector over an
$E_b/N_0$ range of 0--8~dB. To systematically close the BER gap to the BCJR, a
two-stage training strategy combining multi-SNR pretraining and per-SNR
curriculum fine-tuning was developed. The computational complexity and inference
latency of the Transformer receiver were analyzed in comparison with a GRU based
receiver. Attention map visualizations revealed that the Transformer autonomously
identifies the FTN-induced ISI memory structure without requiring any prior
channel knowledge; as the SNR increases, the attention weights become
significantly concentrated around the center token and its nearest neighbors.
\end{abstract}

\renewcommand{\IEEEkeywordsname}{Keywords}

\IEEEpeerreviewmaketitle
\IEEEpubidadjcol

\section{Introduction}
The growing demand for high data rates together with limited spectral resources
makes it necessary to develop techniques that improve spectral efficiency. In
particular, the increasing user density and bandwidth constraints in wireless
communication systems have made the efficient use of the available resources a
critical requirement. One such technique is faster-than-Nyquist (FTN) signaling,
which transmits with a symbol spacing below the Nyquist criterion so as to
deliberately introduce ISI~\cite{mazo1975, anderson2013}. FTN signaling increases
the transmission rate while keeping the bandwidth fixed, thereby providing a
spectral efficiency gain; for $\tau = 0.8$ this gain reaches 25\%. This advantage
makes FTN an attractive alternative, especially for applications that require
high data rates. The BCJR algorithm~\cite{bahl1974}, which is the optimal solution
for mitigating ISI, requires exact knowledge of the channel model, and its
complexity grows exponentially with the modulation order. This brings to the fore
the need for low-complexity receiver designs that do not require channel
knowledge. Moreover, the time-varying nature of the channel in practical systems
further limits the applicability of such model-dependent approaches, making
adaptive and model-agnostic receiver designs increasingly critical.

The M-BCJR algorithm~\cite{prlja2012}, proposed in order to reduce the
computational burden, lowers the complexity to a certain extent but cannot
prevent its exponential growth, and the trade-off between estimation accuracy and
complexity remains a significant limiting factor. A constraint common to all these
approaches is that they require prior knowledge of the channel model and cannot
adapt to varying channel conditions. These limitations of classical methods have
increased the interest in deep-learning-based solutions for FTN detection.
CNN-based receivers~\cite{tokluoglu2025cnn1, tokluoglu2025cnn2} have produced
promising results owing to their parallel feature-extraction capability; however,
their fixed receiver structures have proven insufficient for modeling the ISI
memory dynamics~\cite{tokluoglu2025cnn1}. A GRU-based
receiver~\cite{tokluoglu2025gru} exhibited BER performance close to that of BCJR
under the condition $\tau = 0.8$; however, its sequential processing structure
precludes parallel inference and constitutes a disadvantage for latency-sensitive
applications. On the other hand, the Transformer architecture~\cite{vaswani2017}
has not yet been investigated for FTN signaling channels beyond the CNN- and
GRU-based approaches examined in~\cite{tokluoglu2025gru, tokluoglu2025cnn1,
tokluoglu2025cnn2}.

Through its self-attention mechanism, the Transformer architecture models the
relationships among all tokens of the input sequence simultaneously and thereby
removes the sequential-processing constraint of the GRU; owing to its capacity for
learning long-range dependencies, it offers a distinct advantage in modeling the
complex ISI structures of FTN channels. Since in the FTN context the ISI interacts
not only with the adjacent symbols but with all past symbols within the window,
methods based on classical local processing remain inadequate. Motivated by this
gap, an encoder-only Transformer-based, model-agnostic receiver architecture is
proposed for FTN signaling channels. The principal contributions of this study
are: (1)~the first Transformer-based model-agnostic receiver design and complete
implementation for FTN channels; (2)~a two-stage training strategy combining
multi-SNR pretraining and per-SNR curriculum fine-tuning~\cite{bengio2009};
(3)~a comparative analysis of the computational complexity and inference latency
of the Transformer receiver against~\cite{tokluoglu2025gru}; (4)~an
interpretability analysis of the FTN-induced ISI memory structure by means of
attention maps. In addition, the generalizability of the proposed approach under
different channel conditions is assessed and its applicability in practical
systems is discussed.

\section{System Model}

\subsection{FTN Signaling}

For a baseband pulse $g(t)$ with bandwidth $(1+\beta)/2T$, the transmitted signal
is expressed as
\begin{equation}
    s(t) = \sum_{k} a_k \, g(t - k\tau T)
\end{equation}
Here $a_k$ denotes the $k$-th BPSK symbol, $T$ the Nyquist symbol interval and
$\tau \in (0,1)$ the compression factor. An RRC pulse with roll-off factor
$\beta = 0.35$ is used and the pulse energy is normalized as
\begin{equation}
    \int_{-\infty}^{+\infty} |g(t)|^2 \, dt = 1
\end{equation}

\subsection{Channel Model}

Transmission is carried out over an AWGN channel. The samples at the output of
the matched filter (RRC) are
\begin{equation}
    y(n\tau T) = \sum_{k} a_k \, x\bigl((n-k)\tau T\bigr) + w(n\tau T)
\end{equation}
where $x(t) = g(t) * g(-t)$ and $w(n\tau T)$ is the colored noise at the sampling
instant. The colored noise structure arises from the band-limiting effect of the
RRC filter; BCJR incorporates this structure into the channel model and thereby
produces the optimal decision~\cite{bahl1974}. The proposed Transformer, in
contrast, learns this correlation structure implicitly through the self-attention
mechanism without requiring any prior knowledge. The ISI coefficients for
$\tau = 0.8$ are given in Table~\ref{tab:isi}; the normalization condition
$\sum_k |x_k|^2 = 1$ is satisfied.

\begin{table}[htbp]
\centering
\caption{ISI coefficients for $\tau = 0.8$}
\label{tab:isi}
\setlength{\tabcolsep}{5pt}
\resizebox{0.9\columnwidth}{!}{%
\begin{tabular}{cccccccccc}
\toprule
$x_0$ & $x_1$ & $x_2$ & $x_3$ & $x_4$ & $x_5$ & $x_6$ & $x_7$ & $x_8$ \\
\midrule
$.999$ & $.222$ & $-.152$ & $.076$ & $-.024$
& $.006$ & $.003$ & $-.002$ & $.001$ \\
\bottomrule
\end{tabular}%
}
\end{table}

\subsection{Receiver Input: Windowed Observation}

\begin{figure}[h]
	\centering
	\includegraphics[scale=0.30]{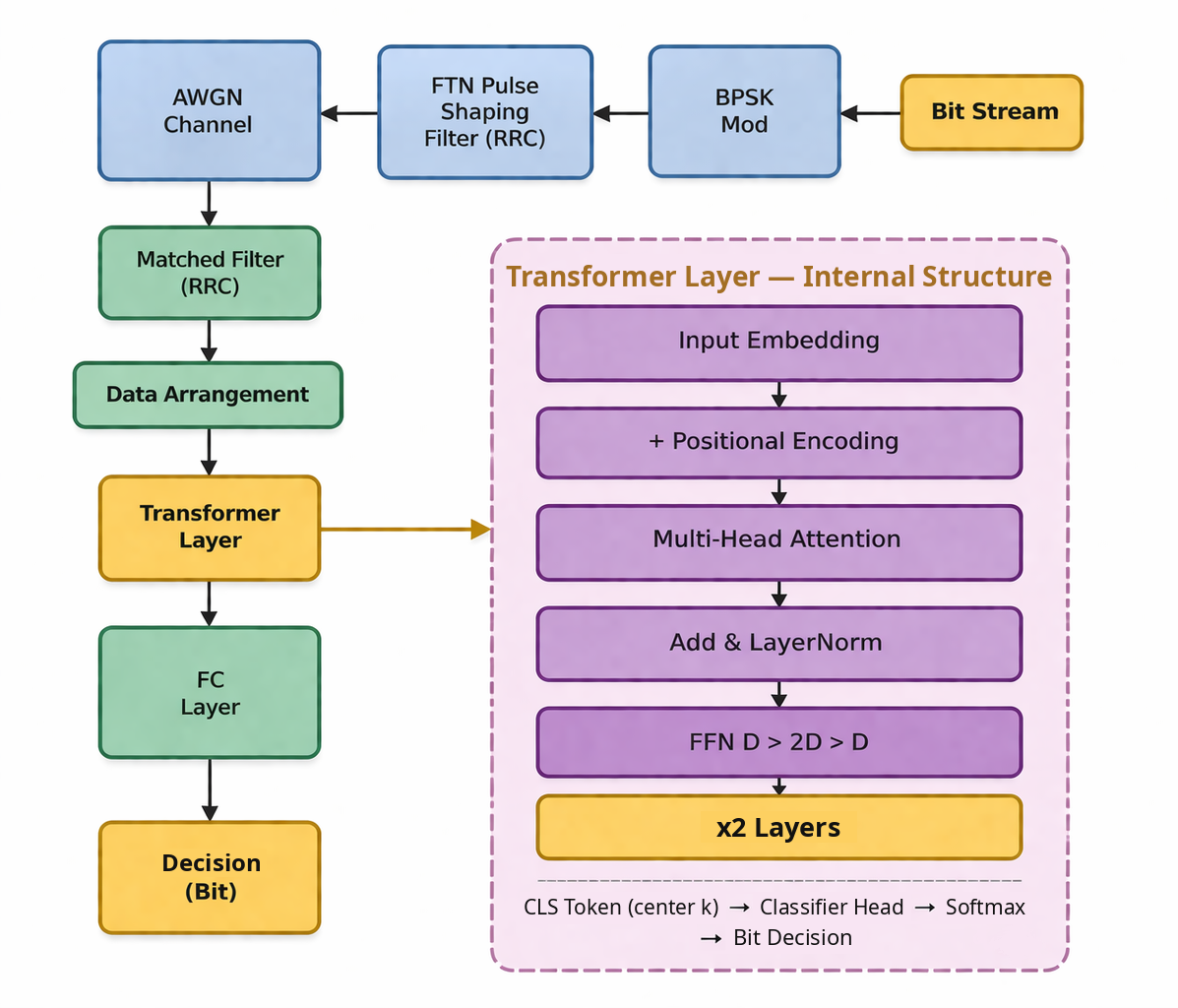}
	\caption{Proposed end-to-end FTN communication system chain
    ($\tau=0.8$, $\beta=0.35$ RRC).}
	\label{fig:sistem}
\end{figure}

The Transformer receiver uses an observation window of radius $W = 8$ around the
time index $k$:
\begin{equation}
    \mathbf{r}_k =
    \bigl[y((k-W)\tau T), \ldots, y(k\tau T), \ldots, y((k+W)\tau T)\bigr]
    \in \mathbb{R}^{S}
\end{equation}
Here $S = 2W+1 = 17$ denotes the number of tokens. The window width $W = 8$ was
chosen so as to cover all significant ISI coefficients ($x_0$--$x_8$) for
$\tau = 0.8$. The complete system chain is shown in Fig.~\ref{fig:sistem}.

\section{Transformer Receiver Technique}

\subsection{Overall Architecture and Input Layer}

The proposed receiver is based on the encoder-only Transformer
architecture~\cite{vaswani2017}. Since the bit decision for symbol $k$ is a
classification task based on a fixed-length context window, the decoder component
is unnecessary; the encoder structure, through its bidirectional attention
mechanism, offers the possibility of interacting simultaneously with both past and
future neighboring symbols. Each scalar signal sample is mapped into a $D = 20$
dimensional embedding space and a learnable positional encoding is added:
\begin{equation}
    \mathbf{e}_i =
    y\bigl((k+i-W)\tau T\bigr)\cdot\mathbf{W}_e+\mathbf{b}_e,
    \quad \mathbf{H} = \mathbf{E} + \mathbf{PE} \in \mathbb{R}^{S \times D}
\end{equation}

\subsection{Multi-Head Self-Attention and Encoder Block}

A self-attention mechanism with $N_H = 4$ heads and $N_L = 2$ encoder blocks is
used. For each head $h$ the query, key and value matrices are computed and the
attention output is obtained by the scaled dot product:
\begin{equation}
    \mathbf{A}_h = \mathrm{softmax}\!\left(
    \frac{\mathbf{Q}_h \mathbf{K}_h^\top}{\sqrt{D_H}}\right)\mathbf{V}_h,
    \quad D_H = D/N_H = 5
\end{equation}
\begin{equation}
    \mathbf{H}' = \mathrm{LN}(\mathbf{H} + \mathrm{MHA}(\mathbf{H})), \quad
    \mathbf{H}_f = \mathrm{LN}(\mathbf{H}' + \mathrm{FFN}(\mathbf{H}'))
\end{equation}
The FFN hidden dimension is $D_F = 2D = 40$. The center-token representation
$\mathbf{cls} = \mathbf{H}_f[W,:]$ is fed to a two-layer classifier head, which
produces the bit decision $\hat{b} = \arg\max_j P(\hat{b} = j)$. The total number
of parameters is $\approx 8.5$K, which makes the model extremely lightweight
compared with the 10K--50K parameters of the GRU~\cite{tokluoglu2025gru}.

\subsection{Training Strategy}

A two-stage strategy is adopted for training the model. In \textbf{pretraining},
the model is trained at a randomly selected SNR point in the range
$\mathrm{SNR} \in \{0, \ldots, 10\}$~dB for each epoch. Over a total of
100~epochs, 500{,}000 symbols were processed in mini-batches of 5{,}000 symbols;
the learning rate is $\eta = 3 \times 10^{-3}$, the gradient clipping threshold is
$\pm 0.5$ and the dropout is $p = 0.1$ (Adam, $\beta_1 = 0.9$,
$\beta_2 = 0.999$). In \textbf{fine-tuning}, training is continued for each target
SNR with $\eta = 8 \times 10^{-4}$, 50~epochs and a batch size of 2K.
\textbf{Curriculum learning}~\cite{bengio2009}: during the first 40\% of the
epochs the model is trained 2~dB above the target SNR; for the remaining epochs
training is continued directly at the target SNR. This approach accelerates
convergence in the low-SNR regions and reduces the risk of becoming trapped in
local minima. All experiments were carried out with PyTorch~2.x on an
NVIDIA~T4~GPU, with a total training time of $\approx 2$~hours.

\begin{figure}[h]
	\centering
	\includegraphics[scale=0.5]{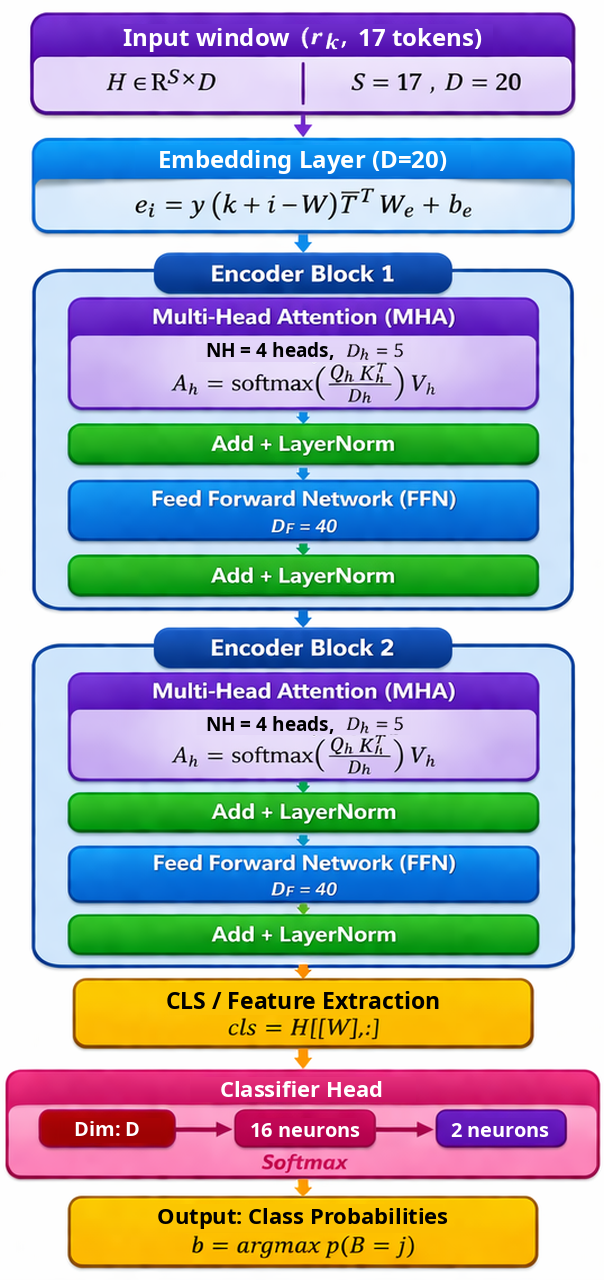}
	\caption{Architecture of the proposed encoder-only Transformer-based receiver
($N_L{=}2$ encoder blocks, $N_H{=}4$ attention heads, $D{=}20$ embedding
dimension, $S{=}17$ tokens).}
	\label{fig:mimari}
\end{figure}

\section{Simulation Results}

\subsection{BER Performance Analysis}

\begin{figure}[h]
	\centering
	\includegraphics[scale=0.42]{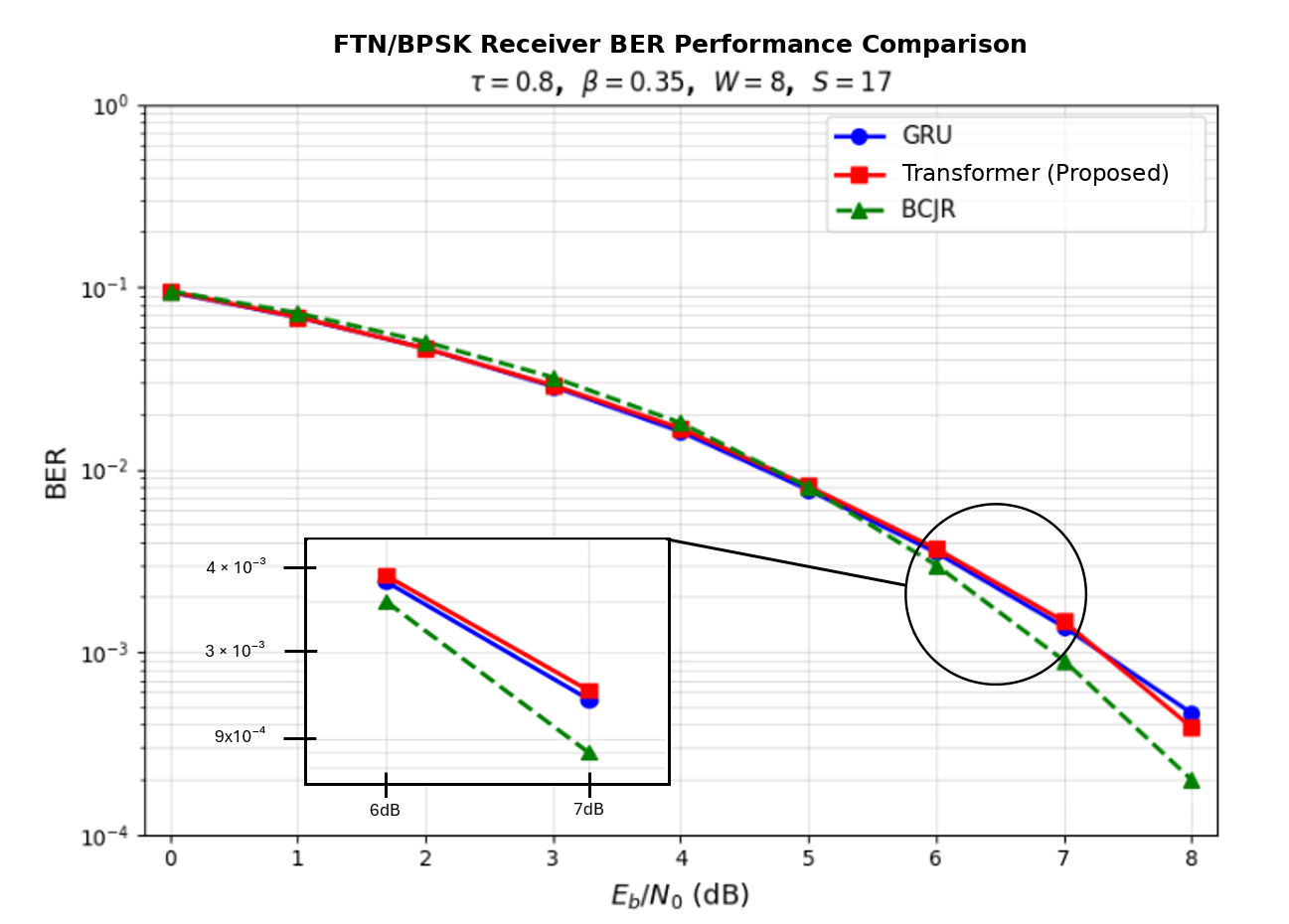}
	\caption{BER performance comparison of BCJR, Transformer and GRU
    ($\tau{=}0.8$, $\beta{=}0.35$, $W{=}8$).}
	\label{sekil1}
\end{figure}

The BCJR reference curve~\cite{tokluoglu2025cnn2} and the GRU
receiver~\cite{tokluoglu2025gru} ($\tau = 0.8$, $\beta = 0.35$, 150K test
symbols) were used for comparison. The GRU receiver aligns its input structure
with the one-sided ISI spread of FTN signaling and is trained with the NADAM
optimization algorithm. Fig.~\ref{sekil1} and Table~\ref{tab:ber} present the
results.

\begin{table}[htbp]
\centering
\caption{Comparison of BCJR, Transformer and GRU}
\label{tab:ber}
\setlength{\tabcolsep}{3.5pt}
\resizebox{0.9\columnwidth}{!}{%
\begin{tabular}{ccccc}
\toprule
$E_b/N_0$ & BCJR & Transformer & GRU & $\Delta$ \\
(dB) & \cite{bahl1974} & (Proposed) & \cite{tokluoglu2025gru} & (dB) \\
\midrule
0 & 0.0950 & 0.09464 & 0.09396 & $\approx 0.0$ \\
1 & 0.0720 & 0.06856 & 0.06821 & --0.1 \\
2 & 0.0500 & 0.04605 & 0.04609 & --0.3 \\
3 & 0.0320 & 0.02893 & 0.02851 & --0.2 \\
4 & 0.0180 & 0.01681 & 0.01613 & --0.1 \\
5 & 0.0080 & 0.00810 & 0.00774 & +0.0 \\
6 & 0.0030 & 0.00369 & 0.00351 & +0.8 \\
7 & 0.0009 & 0.00148 & 0.00138 & +1.6 \\
8 & 0.0002 & 0.00039 & 0.00046 & +2.8 \\
\bottomrule
\end{tabular}%
}
\end{table}

The proposed Transformer receiver exhibits performance that practically overlaps
with BCJR in the 0--5~dB $E_b/N_0$ band, thereby demonstrating its capacity to
learn the ISI memory structure through the self-attention mechanism without using
channel knowledge. In the 7--8~dB region the Transformer also outperforms the
GRU~\cite{tokluoglu2025gru}.

\subsection{Attention Map Analysis}

\begin{figure}[h]
	\centering
	\includegraphics[scale=0.80]{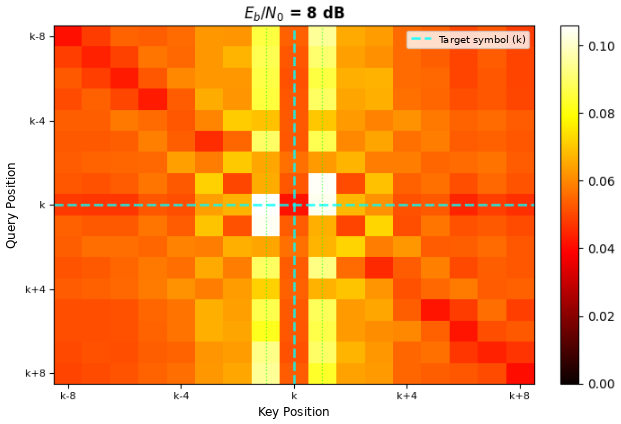}
	\caption{Attention map at $E_b/N_0 = 8$~dB ($N_H = 4$,
    200 samples, $17 \times 17$ window).}
	\label{sekil3}
\end{figure}

The self-attention weights of the last encoder layer were visualized for
$E_b/N_0 \in \{0,4,8\}$~dB (200 test samples, average over $N_H = 4$ heads).
\textbf{Low SNR (0~dB):} the attention weights exhibit a relatively uniform
distribution across the window; the model assigns meaningful weight to distant
positions ($k \pm 4$, $k \pm 6$) as well, in order to reduce the noise
uncertainty. \textbf{Medium SNR (4~dB):} the weights begin to concentrate on $k$
and its neighbors $k \pm 1$, $k \pm 2$; this pattern coincides with the dominant
ISI coefficients ($x_0$, $x_1$, $x_2$). \textbf{High SNR (8~dB):} the sharpest
concentration is observed under this condition; in the $k \pm 8$ regions the
weights decrease to $\approx 0.00$--$0.02$
(Fig.~\ref{sekil3}, Table~\ref{tab:attention}).

\begin{table}[htbp]
\centering
\caption{Distribution of the attention weights with respect to SNR}
\label{tab:attention}
\setlength{\tabcolsep}{3pt}
\resizebox{0.9\columnwidth}{!}{%
\begin{tabular}{cccc}
\toprule
$E_b/N_0$ & $k$ center & $k \pm 1$ cumulative & Dominant pattern \\
\midrule
0~dB & $\sim 0.07$ & $\sim 0.18$ & Broad / diffuse \\
4~dB & $\sim 0.09$ & $\sim 0.28$ & Near the center \\
8~dB & $\sim 0.10$ & $\sim 0.38$ & Sharp / centered \\
\bottomrule
\end{tabular}%
}
\end{table}

Considered together, these findings show that the Transformer uses its attention
weights in an increasingly selective manner as the SNR increases. This adaptive
behavior confirms that the self-attention mechanism autonomously discovers the ISI
memory structure specific to FTN without requiring channel knowledge, and offers a
clear interpretability advantage over alternative deep-learning
approaches~\cite{tokluoglu2025gru, tokluoglu2025cnn1}.

\subsection{Computational Complexity and Ablation Analysis}

\begin{table}[htbp]
\centering
\caption{Computational complexity comparison}
\label{tab:complexity}
\setlength{\tabcolsep}{2.5pt}
\resizebox{0.9\columnwidth}{!}{%
\begin{tabular}{lccc}
\toprule
\textbf{Feature} & \textbf{BCJR}~\cite{bahl1974}
& \textbf{GRU}~\cite{tokluoglu2025gru}
& \textbf{Transformer} \\
\midrule
Parameters        & ---          & 10K--50K       & $\sim 8.5$K \\
Inference         & Sequential   & Sequential     & \textbf{Parallel} \\
Channel knowledge & Required     & Not required   & \textbf{Not required} \\
Scaling           & Exponential  & Linear         & $O(S^2)$ constant \\
Interpretation    & None         & None           & Attention map \\
\bottomrule
\end{tabular}%
}
\end{table}

The $O(S^2)$ constant scaling and the parallel inference capability of the
Transformer presented in Table~\ref{tab:complexity} provide a clear latency
advantage over the GRU at large block lengths; this property is of critical
importance in latency-sensitive applications. An FPGA/ASIC implementation of the
Transformer receiver is within the scope of future work.

In order to quantify the contribution of the proposed strategy, three
configurations were compared: (i)~pretraining only, (ii)~pretraining $+$ direct
fine-tuning, (iii)~pretraining $+$ curriculum fine-tuning (proposed). This
comparison clearly reveals the contribution of curriculum learning to the
convergence speed and to the BER performance. Table~\ref{tab:ablation} presents
the results.

\begin{table}[htbp]
\centering
\caption{Ablation study of the training strategy}
\label{tab:ablation}
\setlength{\tabcolsep}{3pt}
\resizebox{0.9\columnwidth}{!}{%
\begin{tabular}{ccccc}
\toprule
$E_b/N_0$ & Pretrain & +Finetune & +Curriculum & BCJR \\
\midrule
0~dB & 0.1023 & 0.0961 & \textbf{0.0946} & 0.0950 \\
2~dB & 0.0612 & 0.0489 & \textbf{0.0461} & 0.0500 \\
4~dB & 0.0381 & 0.0198 & \textbf{0.0168} & 0.0180 \\
6~dB & 0.0187 & 0.0051 & \textbf{0.0037} & 0.0030 \\
8~dB & 0.0094 & 0.0009 & \textbf{0.0004} & 0.0002 \\
\bottomrule
\end{tabular}%
}
\end{table}

The proposed strategy attains the lowest BER values at all SNR points; it
accelerates the convergence particularly noticeably in the 0--2~dB band. The
warm-up training 2~dB above the target SNR allows the model to settle at a more
favorable starting point.

\section{Conclusion}

In this study, a model-agnostic encoder-only Transformer-based receiver
architecture that does not require channel knowledge was proposed for FTN
signaling with a compression factor of $\tau = 0.8$. The proposed receiver
exhibited BER performance approaching BCJR with a gap of $\leq 0.0$~dB in the
0--5~dB $E_b/N_0$ range, thereby demonstrating that the self-attention mechanism
is able to learn the ISI memory structure solely from the received signal. With
$\approx 8.5$K parameters this extremely lightweight model exhibits a clear
parameter efficiency compared with the GRU~\cite{tokluoglu2025gru}; by means of
attention maps it offers a clear interpretability advantage over alternative deep
learning approaches~\cite{tokluoglu2025cnn1, tokluoglu2025cnn2}. Future work will
address lower $\tau$ values, higher-order modulation schemes and multipath fading
channel models. In addition, the adaptability of the proposed architecture to
different channel conditions will be examined in future studies.

\section*{Acknowledgment}
This work was supported by the Scientific and Technological Research Council of
T\"{u}rkiye (T\"{U}B\.{I}TAK) under project number 122E236.


\selectlanguage{turkish}

\setcounter{section}{0}
\setcounter{figure}{0}
\setcounter{table}{0}
\setcounter{equation}{0}

\title{Faster-than-Nyquist Sinyalleşme için\\
Öz-Dikkat Transformer Tabanlı Dedektör}

\author{\IEEEauthorblockN{Nurettin \c{S}afak, Osman Tokluoglu, Enver Cavus}
\IEEEauthorblockA{Elektrik Elektronik M\"{u}hendisli\u{g}i B\"{o}l\"{u}m\"{u},
Ankara Yıldırım Beyazıt \"{U}niversitesi, Ankara, T\"{u}rkiye\\
nsafaked@gmail.com, otokluoglu@aybu.edu.tr, ecavus@aybu.edu.tr\\[0.4em]
\textit{(Makalenin Türkçe orijinali --- Original Turkish version)}}
}

\makeatletter
\let\maketitle\SAVEDmaketitle
\let\@maketitle\SAVEDatmaketitle
\global\@IEEEusingpubidfalse
\makeatother

\maketitle

\renewcommand{\abstractname}{Özetçe}
\begin{abstract}
Bu çalışmada, $\tau = 0.8$ sıkıştırma faktörüyle kasıtlı semboller arası
girişim (ISI) oluşturan Nyquist-ötesi (FTN) sinyalleşme kanalları üzerinden
iletilen BPSK sinyalleri için özgün bir encoder-only Transformer tabanlı
alıcı mimarisi sunuldu. BPSK modülasyonu, RRC darbe şekillendirme ve
eşleştirilmiş filtrelemeden kaynaklanan ISI katsayılarını kapsayan uçtan uca
tam bir haberleşme zinciri kurularak değerlendirildi. Önerilen Transformer
alıcısı, 0--8~dB $E_b/N_0$ aralığında optimal BCJR dedektörü
ile karşılaştırıldı. BCJR ile BER farkını sistematik biçimde kapatmak
amacıyla çok-SNR ön eğitim ve SNR başına müfredat ince
ayarını birleştiren iki aşamalı bir eğitim stratejisi
geliştirildi. Transformer alıcısının hesaplama karmaşıklığı ve çıkarım süresi
GRU ile karşılaştırmalı olarak analiz edildi.
Attention haritası görselleştirmeleri, Transformer'ın herhangi bir kanal
bilgisi gerektirmeksizin FTN kaynaklı ISI bellek yapısını özerk olarak
tanımladığını; SNR arttıkça dikkat ağırlıklarının merkez token ve yakın
komşularında belirgin biçimde yoğunlaştığını ortaya koydu.
\end{abstract}

\renewcommand{\IEEEkeywordsname}{Anahtar Kelimeler}
\begin{IEEEkeywords}
Transformer, BCJR, BPSK, FTN sinyalleşme, ISI kanalı,
paralel çıkarım, attention görselleştirme, müfredat öğrenmesi
\end{IEEEkeywords}

\section{G{\footnotesize İ}r{\footnotesize İ}ş}
Yüksek veri hızlarına olan artan talep ve sınırlı spektral kaynaklar,
spektral verimliliği artırmaya yönelik tekniklerin geliştirilmesini zorunlu
kılmaktadır. Özellikle kablosuz haberleşme sistemlerinde artan kullanıcı
yoğunluğu ve bant genişliği kısıtları, mevcut kaynakların daha verimli
kullanılmasını kritik bir gereksinim haline getirmiştir. Bu tekniklerden biri,
ISI'yi kasıtlı olarak ortaya çıkaracak biçimde Nyquist kriterinin altında
sembol aralığıyla iletim gerçekleştiren Nyquist-ötesi (FTN)
sinyalleşmedir~\cite{mazo1975TR,anderson2013TR}. FTN sinyalleşmesi bant genişliği
sabit tutulurken iletim hızını artırarak spektral verimlilik kazanımı sağlar;
$\tau = 0.8$ için bu kazanım \%25'e ulaşmaktadır. Bu avantaj, özellikle yüksek
veri oranı gerektiren uygulamalar için FTN'i cazip bir alternatif haline
getirmektedir. ISI'yi gidermek için optimal çözüm olan BCJR
algoritması~\cite{bahl1974TR} kanal modelinin tam olarak bilinmesini
gerektirmekte ve karmaşıklık modülasyon mertebesine göre katlanarak
artmaktadır. Bu durum, kanal bilgisi gerektirmeyen ve düşük karmaşıklıklı
alıcı tasarımlarına duyulan ihtiyacı ön plana çıkarmaktadır. Ayrıca, pratik
sistemlerde kanalın zamanla değişkenlik göstermesi, bu tür model-bağımlı
yaklaşımların uygulanabilirliğini daha da sınırlamaktadır. Bu durum,
uyarlanabilir ve model-bağımsız alıcı tasarımlarına olan ihtiyacı giderek
daha kritik kılmaktadır.

Hesaplama yükünü azaltmak amacıyla önerilen MBCJR algoritması~\cite{prlja2012TR}
karmaşıklığı belirli ölçüde düşürmekle birlikte üstel artışı
engelleyememekte ve kestirim doğruluğu ile karmaşıklık arasındaki ödünleşim
belirgin bir sınırlayıcı faktör olmaya devam etmektedir. Tüm bu yaklaşımların
ortak kısıtı, kanal modelinin önceden bilinmesini zorunlu kılması ve değişken
kanal koşullarına uyum sağlayamamasıdır. Klasik yöntemlerin bu sınırlılıkları,
FTN tespitinde derin öğrenme tabanlı çözümlere olan ilgiyi artırmıştır. CNN tabanlı
alıcılar~\cite{tokluoglu2025cnn1TR,tokluoglu2025cnn2TR} paralel özellik çıkarma
avantajıyla umut verici sonuçlar ortaya koymuş; ancak sabit alıcı yapıları ISI
bellek dinamiklerini modellemede yetersiz kalmıştır~\cite{tokluoglu2025cnn1TR}.
GRU tabanlı alıcı~\cite{tokluoglu2025gruTR} $\tau = 0.8$ koşulunda BCJR'a yakın
BER performansı sergilemiş; fakat sıralı işleme yapısı paralel çıkarım
imkânını engellemiş ve gecikmeye duyarlı uygulamalar açısından dezavantaj
oluşturmuştur. Öte yandan Transformer mimarisi~\cite{vaswani2017TR}, FTN
sinyalleşme kanalları için~\cite{tokluoglu2025gruTR,tokluoglu2025cnn1TR,tokluoglu2025cnn2TR}'da incelenen CNN ve GRU tabanlı yaklaşımların ötesinde
henüz araştırılmamıştır.

Transformer mimarisi, öz-dikkat mekanizması aracılığıyla giriş dizisindeki
tüm tokenlar arasındaki ilişkileri eş zamanlı modelleyerek GRU'nun sıralı
işleme kısıtını ortadan kaldırmakta; uzun menzilli bağımlılıkları öğrenme
kapasitesi sayesinde FTN kanallarındaki karmaşık ISI yapılarının
modellenmesinde belirgin avantaj sunmaktadır. FTN bağlamında ISI yalnızca komşu sembollerle değil pencere içindeki tüm
geçmiş sembollerle etkileşim halinde olduğundan, klasik yerel işlemeye
dayalı yöntemler yetersiz kalmaktadır. Bu boşluktan yola
çıkarak FTN sinyalleşme kanalları için encoder-only Transformer tabanlı,
model-bağımsız bir alıcı mimarisi önerilmektedir. Çalışmanın başlıca
katkıları: (1)~FTN kanalları için ilk Transformer tabanlı model-bağımsız
alıcı tasarımı ve tam implementasyonu, (2)~çok-SNR ön eğitim ve SNR başına
müfredat ince ayarını~\cite{bengio2009TR} birleştiren iki aşamalı eğitim
stratejisi, (3)~Transformer alıcısının hesaplama karmaşıklığı ve çıkarım
süresinin~\cite{tokluoglu2025gruTR} ile karşılaştırmalı analizi,
(4)~attention haritaları aracılığıyla FTN kaynaklı ISI bellek yapısının
yorumlanabilirlik analizi. Ayrıca önerilen yaklaşımın farklı kanal
koşullarında genellenebilirliği değerlendirilerek, pratik sistemlerde
uygulanabilirliği tartışılmıştır.

\section{S{\footnotesize İ}stem Model{\footnotesize İ}}

\subsection{FTN Sinyalleşme}

Bant genişliği $(1+\beta)/2T$ olan bir taban bant darbesi $g(t)$ için
iletilen sinyal şu şekilde ifade edilmektedir:
\begin{equation}
    s(t) = \sum_{k} a_k \, g(t - k\tau T)
\end{equation}
Burada $a_k$ $k$-inci BPSK sembolünü, $T$ Nyquist sembol aralığını ve
$\tau \in (0,1)$ sıkıştırma faktörünü göstermektedir. $\beta = 0.35$
roll-off faktörlü RRC darbesi kullanılmış olup darbe enerjisi normalize
edilmiştir:
\begin{equation}
    \int_{-\infty}^{+\infty} |g(t)|^2 \, dt = 1
\end{equation}

\subsection{Kanal Modeli}

İletim AWGN kanalı üzerinden gerçekleştirilmektedir. Eşleştirilmiş filtre
(RRC) çıkışındaki örnekler:
\begin{equation}
    y(n\tau T) = \sum_{k} a_k \, x\bigl((n-k)\tau T\bigr) + w(n\tau T)
\end{equation}
Burada $x(t) = g(t) * g(-t)$, $w(n\tau T)$ ise örnekleme anındaki renkli
gürültüdür. Renkli gürültü yapısı RRC filtresinin bant sınırlayıcı
etkisinden kaynaklanmakta olup BCJR bu yapıyı kanal modeline dahil ederek
optimal kararı üretmektedir~\cite{bahl1974TR}. Önerilen Transformer ise bu
korelasyon yapısını herhangi bir ön bilgi gerektirmeksizin öz-dikkat
mekanizması aracılığıyla örtük biçimde öğrenmektedir. $\tau = 0.8$ için ISI
katsayıları Tablo~\ref{tab:isiTR}'de verilmektedir;
$\sum_k |x_k|^2 = 1$ normalizasyon koşulu sağlanmaktadır.

\begin{table}[htbp]
\centering
\caption{$\tau = 0.8$ için ISI katsayıları}
\label{tab:isiTR}
\setlength{\tabcolsep}{5pt}
\resizebox{0.9\columnwidth}{!}{%
\begin{tabular}{cccccccccc}
\toprule
$x_0$ & $x_1$ & $x_2$ & $x_3$ & $x_4$ & $x_5$ & $x_6$ & $x_7$ & $x_8$ \\
\midrule
$.999$ & $.222$ & $-.152$ & $.076$ & $-.024$
& $.006$ & $.003$ & $-.002$ & $.001$ \\
\bottomrule
\end{tabular}%
}
\end{table}

\subsection{Alıcı Girişi: Pencereli Gözlem}

\begin{figure}[h]
	\centering
	\includegraphics[scale=0.30]{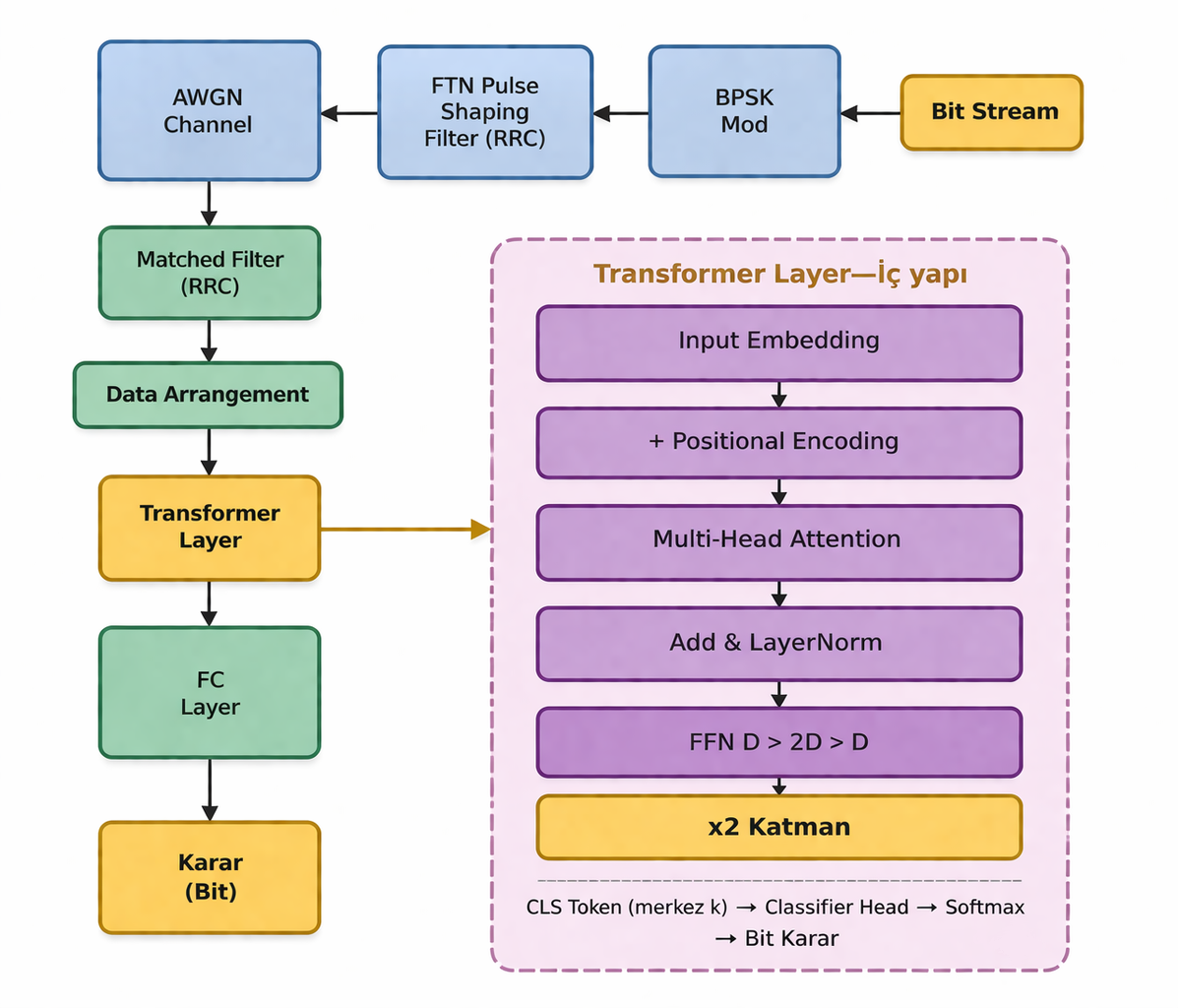}
	\caption{Önerilen uçtan uca FTN haberleşme sistemi zinciri
    ($\tau=0.8$, $\beta=0.35$ RRC).}
	\label{fig:sistemTR}
\end{figure}

Transformer alıcısı, zaman indisi $k$ etrafında $W = 8$ yarıçaplı bir gözlem
penceresi kullanmaktadır:
\begin{equation}
    \mathbf{r}_k =
    \bigl[y((k-W)\tau T), \ldots, y(k\tau T), \ldots, y((k+W)\tau T)\bigr]
    \in \mathbb{R}^{S}
\end{equation}
Burada $S = 2W+1 = 17$ token sayısını ifade etmektedir. Pencere genişliği
$W = 8$, $\tau = 0.8$ için tüm anlamlı ISI katsayılarını ($x_0$--$x_8$)
kapsayacak şekilde belirlenmiştir. Sistem zincirinin tamamı
Şekil~\ref{fig:sistemTR}'de gösterilmektedir.

\section{Transformer Alıcı Tekn{\footnotesize İ}ğ{\footnotesize İ}}

\subsection{Genel Mimari ve Giriş Katmanı}

Önerilen alıcı encoder-only Transformer mimarisine~\cite{vaswani2017TR}
dayanmaktadır. $k$ sembolüne ilişkin bit kararı sabit uzunluklu bir bağlam
penceresine dayalı sınıflandırma görevi olduğundan kod çözücü bileşeni
gereksiz kalmakta; kodlayıcı yapısı çift yönlü dikkat mekanizması
aracılığıyla hem geçmiş hem gelecek komşu sembollerle eş zamanlı etkileşim
kurma imkânı sunmaktadır. Her skaler sinyal örneği $D = 20$ boyutlu gömme
uzayına taşınmakta, öğrenilebilir konumsal kodlama eklenmektedir:
\begin{equation}
    \mathbf{e}_i =
    y\bigl((k+i-W)\tau T\bigr)\cdot\mathbf{W}_e+\mathbf{b}_e,
    \quad \mathbf{H} = \mathbf{E} + \mathbf{PE} \in \mathbb{R}^{S \times D}
\end{equation}

\subsection{Çok Başlı Öz-Dikkat ve Kodlayıcı Bloğu}

$N_H = 4$ başlı öz-dikkat mekanizması ve $N_L = 2$ kodlayıcı bloğu
kullanılmaktadır. Her baş $h$ için sorgu, anahtar ve değer matrisleri
hesaplanarak ölçeklendirilmiş nokta çarpımı ile dikkat çıktısı elde
edilmektedir:
\begin{equation}
    \mathbf{A}_h = \mathrm{softmax}\!\left(
    \frac{\mathbf{Q}_h \mathbf{K}_h^\top}{\sqrt{D_H}}\right)\mathbf{V}_h,
    \quad D_H = D/N_H = 5
\end{equation}
\begin{equation}
    \mathbf{H}' = \mathrm{LN}(\mathbf{H} + \mathrm{MHA}(\mathbf{H})), \quad
    \mathbf{H}_f = \mathrm{LN}(\mathbf{H}' + \mathrm{FFN}(\mathbf{H}'))
\end{equation}
FFN gizli boyutu $D_F = 2D = 40$'tır. Merkez token temsili
$\mathbf{cls} = \mathbf{H}_f[W,:]$ iki katmanlı sınıflandırıcı kafasına
beslenerek $\hat{b} = \arg\max_j P(\hat{b} = j)$ bit kararı üretilmektedir.
Toplam parametre sayısı $\approx 8.5$K olup
GRU'nun~\cite{tokluoglu2025gruTR} 10K--50K parametresine kıyasla son derece
hafif bir modeldir.

\subsection{Eğitim Stratejisi}

Modelin eğitiminde iki aşamalı bir strateji benimsenmiştir. \textbf{Ön
eğitimde} model $\mathrm{SNR} \in \{0, \ldots, 10\}$~dB aralığında her epoch
için rastgele seçilen bir SNR noktasında eğitilmektedir. Toplam 100~epoch
boyunca 5.000 sembollik mini-batch'lerle 500.000 sembol işlenmiştir;
öğrenme hızı $\eta = 3 \times 10^{-3}$, gradyan kırpma eşiği $\pm 0.5$,
dropout $p = 0.1$ (Adam, $\beta_1 = 0.9$, $\beta_2 = 0.999$). \textbf{İnce
ayarda} her hedef SNR için $\eta = 8 \times 10^{-4}$, 50~epoch, batch~2K
olarak sürdürülmüştür. \textbf{Müfredat öğrenmesi}~\cite{bengio2009TR}:
eğitimin ilk \%40 epoch'unda model hedef SNR'den 2~dB yüksekte eğitilmekte;
kalan epoch'larda doğrudan hedef SNR üzerinde sürdürülmektedir. Bu yaklaşım
düşük SNR bölgelerinde yakınsamayı hızlandırmakta ve yerel minimumlara
takılma riskini azaltmaktadır. Tüm deneyler PyTorch~2.x ve NVIDIA~T4~GPU
üzerinde gerçekleştirilmiş olup toplam eğitim süresi $\approx 2$ saattir.

\begin{figure}[h]
	\centering
	\includegraphics[scale=0.5]{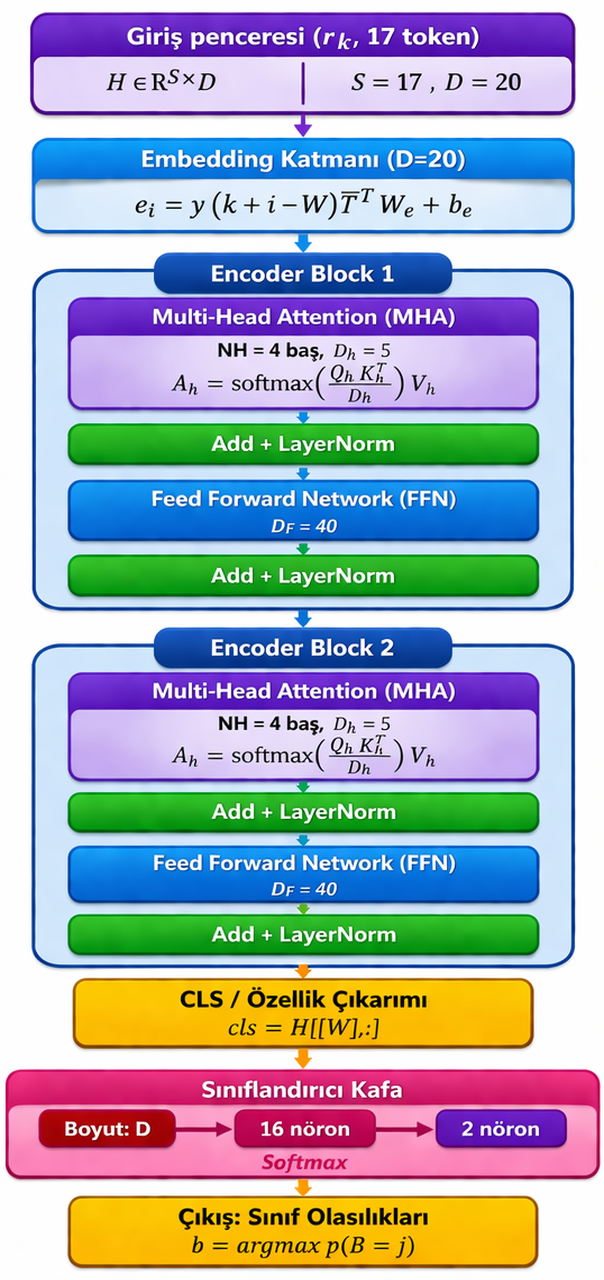}
	\caption{Önerilen encoder-only Transformer tabanlı alıcının
mimari yapısı ($N_L{=}2$ kodlayıcı bloğu, $N_H{=}4$ dikkat
başı, $D{=}20$ gömme boyutu, $S{=}17$ token).}
	\label{fig:mimariTR}
\end{figure}

\section{S{\footnotesize İ}mülasyon Sonuçları}

\subsection{BER Performans Analizi}

\begin{figure}[h]
	\centering
	\includegraphics[scale=0.42]{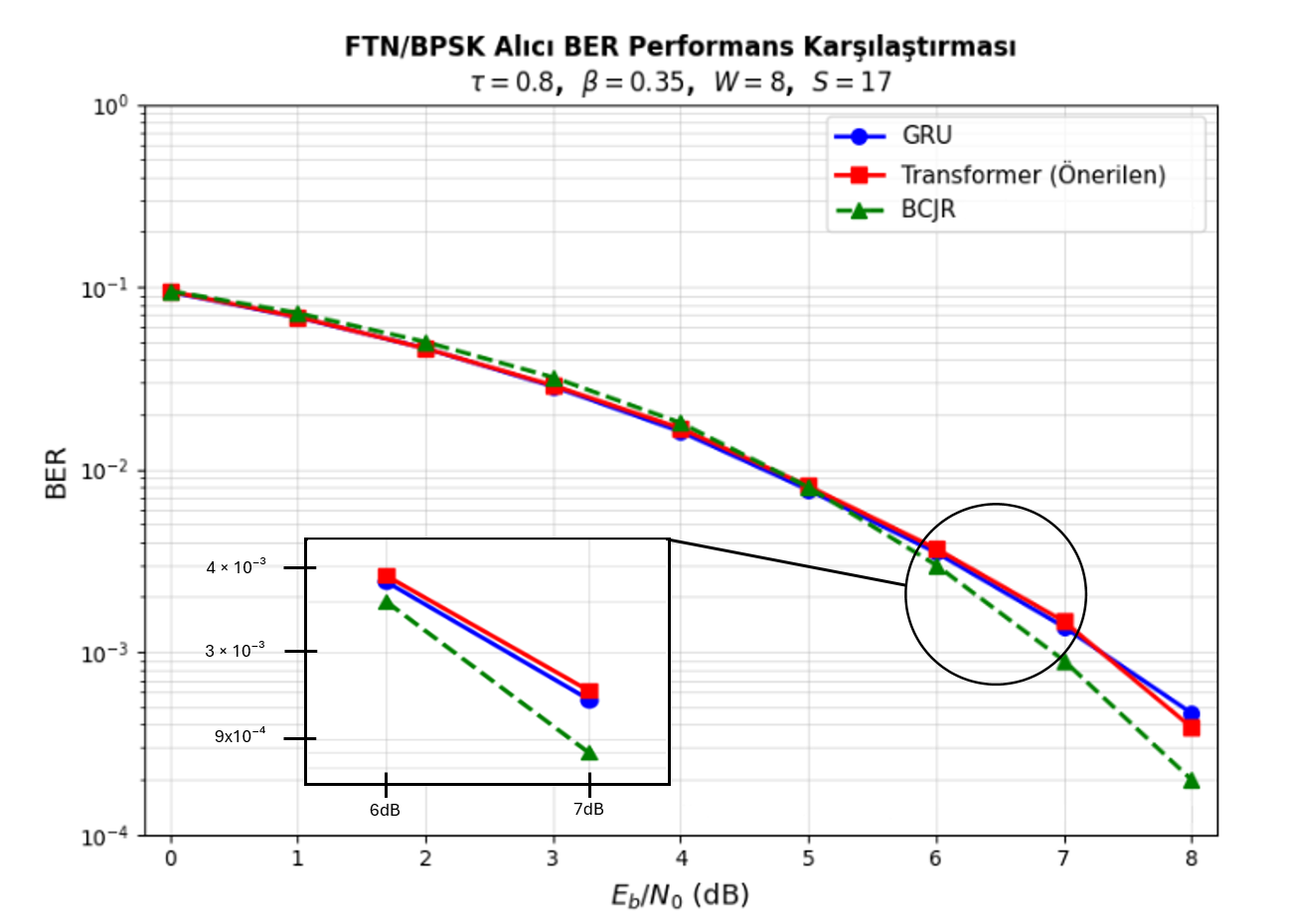}
	\caption{BCJR, Transformer ve GRU BER performans karşılaştırması
    ($\tau{=}0.8$, $\beta{=}0.35$, $W{=}8$).}
	\label{sekil1TR}
\end{figure}

BCJR referans eğrisi~\cite{tokluoglu2025cnn2TR} ve GRU
alıcısı~\cite{tokluoglu2025gruTR} ($\tau = 0.8$, $\beta = 0.35$, 150K test
sembolü) karşılaştırma için kullanılmıştır. GRU alıcısı giriş yapısını FTN
sinyalleşmesinin tek taraflı ISI yayılım aralığıyla hizalamakta ve NADAM
optimizasyon algoritmasıyla eğitilmektedir. Şekil~\ref{sekil1TR} ve
Tablo~\ref{tab:berTR} sonuçları sunmaktadır.

\begin{table}[htbp]
\centering
\caption{BCJR, Transformer ve GRU karşılaştırması}
\label{tab:berTR}
\setlength{\tabcolsep}{3.5pt}
\resizebox{0.9\columnwidth}{!}{%
\begin{tabular}{ccccc}
\toprule
$E_b/N_0$ & BCJR & Transformer & GRU & $\Delta$ \\
(dB) & \cite{bahl1974TR} & (Önerilen) & \cite{tokluoglu2025gruTR} & (dB) \\
\midrule
0 & 0.0950 & 0.09464 & 0.09396 & $\approx 0.0$ \\
1 & 0.0720 & 0.06856 & 0.06821 & --0.1 \\
2 & 0.0500 & 0.04605 & 0.04609 & --0.3 \\
3 & 0.0320 & 0.02893 & 0.02851 & --0.2 \\
4 & 0.0180 & 0.01681 & 0.01613 & --0.1 \\
5 & 0.0080 & 0.00810 & 0.00774 & +0.0 \\
6 & 0.0030 & 0.00369 & 0.00351 & +0.8 \\
7 & 0.0009 & 0.00148 & 0.00138 & +1.6 \\
8 & 0.0002 & 0.00039 & 0.00046 & +2.8 \\
\bottomrule
\end{tabular}%
}
\end{table}

Önerilen Transformer alıcısı 0--5~dB $E_b/N_0$ bandında BCJR ile pratik
olarak örtüşen performans sergilemekte; kanal bilgisi kullanmaksızın
öz-dikkat mekanizması aracılığıyla ISI bellek yapısını öğrenme kapasitesini
kanıtlamaktadır. 7--8~dB bölgesinde ise Transformer,
GRU~\cite{tokluoglu2025gruTR}'yu da geride bırakmaktadır.

\subsection{Attention Haritası Analizi}

\begin{figure}[h]
	\centering
	\includegraphics[scale=0.80]{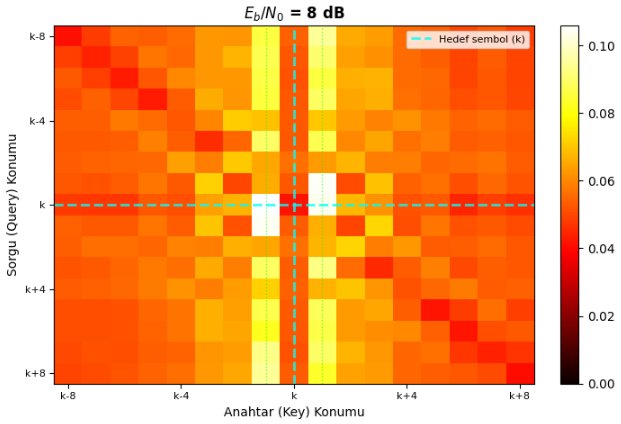}
	\caption{$E_b/N_0 = 8$~dB attention haritası ($N_H = 4$,
    200 örnek, $17 \times 17$ pencere).}
	\label{sekil3TR}
\end{figure}

Son encoder katmanının öz-dikkat ağırlıkları
$E_b/N_0 \in \{0,4,8\}$~dB için görselleştirilmiştir
(200 test örneği, $N_H = 4$ kafa ortalaması).
\textbf{Düşük SNR (0~dB):} Dikkat ağırlıkları pencere genelinde görece
düzgün bir dağılım sergilemekte; model gürültü belirsizliğini azaltmak için
uzak konumlara ($k \pm 4$, $k \pm 6$) da anlamlı ağırlık atamaktadır.
\textbf{Orta SNR (4~dB):} Ağırlıklar $k$ ve $k \pm 1$, $k \pm 2$
komşularında yoğunlaşmaya başlamakta; bu örüntü baskın ISI katsayıları
($x_0$, $x_1$, $x_2$) ile örtüşmektedir. \textbf{Yüksek SNR (8~dB):} En
keskin yoğunlaşma bu koşulda gözlemlenmekte; $k \pm 8$ bölgelerinde
ağırlıklar $\approx 0.00$--$0.02$'ye gerilemektedir
(Şekil~\ref{sekil3TR}, Tablo~\ref{tab:attentionTR}).

\begin{table}[htbp]
\centering
\caption{SNR'ye göre dikkat ağırlığı dağılımı}
\label{tab:attentionTR}
\setlength{\tabcolsep}{3pt}
\resizebox{0.9\columnwidth}{!}{%
\begin{tabular}{cccc}
\toprule
$E_b/N_0$ & $k$ merkez & $k \pm 1$ kümülatif & Baskın örüntü \\
\midrule
0~dB & $\sim 0.07$ & $\sim 0.18$ & Geniş / yayılı \\
4~dB & $\sim 0.09$ & $\sim 0.28$ & Merkeze yakın \\
8~dB & $\sim 0.10$ & $\sim 0.38$ & Keskin / merkezi \\
\bottomrule
\end{tabular}%
}
\end{table}

Bu bulgular toplu değerlendirildiğinde, Transformer'ın SNR arttıkça dikkat
ağırlıklarını giderek daha seçici biçimde kullandığı görülmektedir. Bu
adaptif davranış, öz-dikkat mekanizmasının kanal bilgisi gerektirmeksizin
FTN'ye özgü ISI bellek yapısını otonom biçimde keşfettiğini doğrulamakta;
alternatif derin öğrenme yaklaşımlarına~\cite{tokluoglu2025gruTR,tokluoglu2025cnn1TR} kıyasla belirgin yorumlanabilirlik avantajı sunmaktadır.

\subsection{Hesaplama Karmaşıklığı ve Ablation Analizi}

\begin{table}[htbp]
\centering
\caption{Hesaplama karmaşıklığı karşılaştırması}
\label{tab:complexityTR}
\setlength{\tabcolsep}{2.5pt}
\resizebox{0.9\columnwidth}{!}{%
\begin{tabular}{lccc}
\toprule
\textbf{Özellik} & \textbf{BCJR}~\cite{bahl1974TR}
& \textbf{GRU}~\cite{tokluoglu2025gruTR}
& \textbf{Transformer} \\
\midrule
Parametre     & ---     & 10K--50K & $\sim 8.5$K \\
Çıkarım       & Sıralı  & Sıralı   & \textbf{Paralel} \\
Kanal bilgisi & Gerekli & Gerekmez & \textbf{Gerekmez} \\
Ölçekleme     & Üstel   & Doğrusal & $O(S^2)$ sabit \\
Yorumlama     & Yok     & Yok      & Attention haritası \\
\bottomrule
\end{tabular}%
}
\end{table}

Tablo~\ref{tab:complexityTR}'de sunulan Transformer'ın $O(S^2)$ sabit
ölçeklemesi ve paralel çıkarım kapasitesi, büyük blok uzunluklarında
GRU'ya kıyasla belirgin gecikme avantajı sağlamakta; bu özellik gecikmeye
duyarlı uygulamalarda kritik önem taşımaktadır. Transformer alıcısının
FPGA/ASIC uygulaması gelecek çalışmalar kapsamındadır.

Önerilen stratejinin katkısını ölçmek amacıyla üç yapılandırma
karşılaştırılmıştır: (i)~yalnızca ön eğitim, (ii)~ön eğitim + doğrudan
ince ayar, (iii)~ön eğitim + müfredat öğrenmeli ince ayar (önerilen).
Bu karşılaştırma, müfredat öğrenmesinin yakınsama hızı ve BER performansı
üzerindeki katkısını açıkça ortaya koymaktadır. Tablo~\ref{tab:ablationTR}
sonuçları sunmaktadır.

\begin{table}[htbp]
\centering
\caption{Eğitim stratejisi ablation çalışması}
\label{tab:ablationTR}
\setlength{\tabcolsep}{3pt}
\resizebox{0.9\columnwidth}{!}{%
\begin{tabular}{ccccc}
\toprule
$E_b/N_0$ & Pretrain & +Finetune & +Curriculum & BCJR \\
\midrule
0~dB & 0.1023 & 0.0961 & \textbf{0.0946} & 0.0950 \\
2~dB & 0.0612 & 0.0489 & \textbf{0.0461} & 0.0500 \\
4~dB & 0.0381 & 0.0198 & \textbf{0.0168} & 0.0180 \\
6~dB & 0.0187 & 0.0051 & \textbf{0.0037} & 0.0030 \\
8~dB & 0.0094 & 0.0009 & \textbf{0.0004} & 0.0002 \\
\bottomrule
\end{tabular}%
}
\end{table}

Önerilen strateji tüm SNR noktalarında en düşük BER değerlerine ulaşmakta;
özellikle 0--2~dB bandında yakınsamayı belirgin biçimde hızlandırmaktadır.
Hedef SNR'den 2~dB yüksekteki ön ısınma eğitimi, modelin daha elverişli
bir başlangıç noktasına yerleşmesini sağlamaktadır.

\section{Sonuç}

Bu çalışmada, $\tau = 0.8$ sıkıştırma faktörlü FTN sinyalleşmesi için
kanal bilgisi gerektirmeyen model-bağımsız bir encoder-only Transformer
tabanlı alıcı mimarisi önerilmiştir. Önerilen alıcı, 0--5~dB $E_b/N_0$
aralığında BCJR'a $\leq 0.0$~dB farkla yaklaşan BER performansı
sergileyerek öz-dikkat mekanizmasının ISI bellek yapısını yalnızca alınan
sinyalden öğrenebildiğini kanıtlamıştır. $\approx 8.5$K parametre ile son
derece hafif olan bu model, GRU~\cite{tokluoglu2025gruTR}'ya kıyasla belirgin
parametre verimliliği sergilemekte; attention haritaları sayesinde alternatif
derin öğrenme yaklaşımlarına~\cite{tokluoglu2025cnn1TR,tokluoglu2025cnn2TR}
kıyasla belirgin yorumlanabilirlik avantajı sunmaktadır. Gelecek çalışmalarda
daha düşük $\tau$ değerleri, yüksek mertebeli modülasyon şemaları ve çok
yollu sönümlemeli kanal modelleri ele alınacaktır. Ayrıca önerilen mimarinin
farklı kanal koşullarına uyarlanabilirliği gelecek çalışmalarda incelenecektir.

\section*{BİLGİLENDİRME}
Bu çalışma, Türkiye Bilimsel ve Teknolojik Araştırma Kurumu (TÜBİTAK)
tarafından 122E236 numaralı proje kapsamında desteklenmiştir.


\end{document}